\documentclass{elsart}

\usepackage{graphics}
\usepackage{graphicx}
\usepackage{epsfig}

\usepackage{amssymb}

\begin{document}

\begin{frontmatter}



\title{Evolutionary game dynamics with three strategies
in finite populations}


\author[ad1,ad2]{Jing Wang},
\author[ad1,ad2]{Feng Fu},
\author[ad1,ad2]{Long Wang\corauthref{cor}},
\ead{longwang@pku.edu.cn} \corauth[cor]{Corresponding author. Fax:
+86-010-6275-4388.}
\author[ad1,ad2]{Guangming Xie}

\address[ad1]{Intelligent Control Laboratory, Center for Systems and Control, Department of
Mechanics and Space Technologies, College of Engineering, Peking
University, Beijing 100871, China }

\address[ad2]{Department of Industrial
Engineering and Management, College of Engineering, Peking
University, Beijing 100871, China}

\begin{abstract}
We propose a model for evolutionary game dynamics with three
strategies $A$, $B$ and $C$ in the framework of Moran process in
finite populations. The model can be described as a stochastic
process which can be numerically computed from a system of linear
equations. Furthermore, to capture the feature of the evolutionary
process, we define two essential variables, the {\em global} and the
{\em local} fixation probability. If the {\em global} fixation
probability of strategy $A$ exceeds the neutral fixation
probability, the selection favors $A$ replacing $B$ or $C$ no matter
what the initial ratio of $B$ to $C$ is. Similarly, if the {\em
local} fixation probability of $A$ exceeds the neutral one, the
selection favors $A$ replacing $B$ or $C$ only in some appropriate
initial ratios of $B$ to $C$. Besides, using our model, the famous
game with AllC, AllD and TFT is analyzed. Meanwhile, we find that a
single individual TFT could invade the entire population under
proper conditions.
\end{abstract}

\end{frontmatter}


\section{Introduction}
\label{}

Since the theory of games was first explicitly applied in
evolutionary biology by Lewontin \cite{1}, it has undergone
extensive development. Evolutionary game theory as a mathematical
framework to depict the population dynamics under natural selection
has attracted much attention for a long time. It is a way of
thinking about particular phenotypes depending on their frequencies
in the population \cite{2}. Much more important results in terms of
the corresponding replicator dynamics \cite{3,4}, in which the size
of the well-mixed population is infinite, promote the development of
the evolutionary game theory. Replicator dynamics, which is due to
Taylor and Jonker \cite{5}, is a system of deterministic
differential equations which could describe the evolutionary
dynamics of multi-species. Besides replicator dynamics, the
Lotka-Volterra equations which were devised by Lotka and Volterra,
have received much attention. They are the most common models for
the dynamics of the population numbers \cite{3,6}, whereas
replicator dynamics is the most common model for the evolution of
the frequencies of strategies in a population. These two
deterministic models both fail to account for stochastic effects.
Thus the theory of stochastic processes plays an extraordinarily
important role in
depicting the evolutionary dynamics. \\
In nature, however, populations are finite in size. Finite
population effects can be neglected in infinite populations, but
affect the evolution in finite size. In fact, with high probability,
the state of the process for large size $N$ remains close to the
results of corresponding deterministic replicator dynamics for some
large time $T$ \cite{7}. Recently, an explicit mean-field
description in the form of Fokker-Planck equation was derived for
frequency-dependent selection in finite populations \cite{8,9,10}.
It is an approach which could connect the situation of finite
populations with that of infinite populations. The explicit rules
governing the interaction of a finite number of individuals with
each other are embodied in a master equation. And the finite size
effects are captured in the drift and diffusion terms of a
Fokker-Planck equation where the diffusion term vanishes with $1/N$
for increasing population sizes. This framework was extended to an
evolutionary game with an arbitrary number of strategies \cite{11}.
The stochastic evolutionary processes would be characterized by the
stochastic replicator-mutator equation
$\dot{x_{i}}=a_{i}(\vec{x})+\sum_{j=1}^{d-1}c_{ij}(\vec{x})\xi_{j}(t)$,
here $x_{i}$ means the density of the $i$th individuals using one of
the arbitrary $d$ strategies. Note that the difference between above
equation and the replicator dynamics in infinite populations is the
uncorrelated Gaussian noises term
$\sum_{j=1}^{d-1}c_{ij}(\vec{x})\xi_{j}(t)$. Thus the finite size
effects can be viewed as combinations of some uncorrelated Gaussian
noises. \\
Nowak introduced the frequency-dependent Moran process into
evolutionary game in finite populations \cite{12,13,14}. The
frequency-dependent Moran process is a stochastic birth-death
process. It follows two steps: selection, a player is selected to
reproduce with a probability proportional to its fitness, and the
offspring will use the same strategy as its parent; replacement, a
randomly selected individual is replaced by the offspring. Hence,
the population size, $N$, is strictly constant \cite{15}. Suppose a
population consists of individuals who use either strategy $A$ or
$B$. Individual using strategy $A$ receives payoff $a$ or $b$
playing with $A$ or $B$ individual; individual using strategy $B$
obtains payoff $c$ or $d$ playing with $A$ or $B$ individual.
Viewing individuals using strategy $A$ as mutants, we get the
probability $\bar{x}(i)$ that $i$ mutants could invade and take over
the whole populations. The fitness of individuals using strategy $A$
and $B$ is respectively given by:
\begin{equation}\label{e1}
\begin{array}{l}
f_i=1-w+w[a(i-1)+b(N-i)]/(N-1)\\
g_i=1-w+w[ci+d(N-i-1)]/(N-1)\\
\end{array}
\end{equation}
Here $i=1,2,\cdots,N-1$, $w \in[0,1]$ describes the contribution of
the game to the fitness.   For neutral selection, it needs $w=0$;
for weak selection, it must satisfy the condition $w\ll1$.
Accordingly, the fixation probability $\bar{x}(i)$ is given by
\cite{16}:
\begin{equation}\label{e2}
\begin{array}{llll}
\bar{x}(0)&=&0, \bar{x}(N)=1& \\
\bar{x}(i)&=&\frac{\displaystyle 1+\sum_{j=1}^{i-1}\prod_{k=1}^{j}\frac{g_k}{f_k}}{\displaystyle 1+\sum_{j=1}^{N-1}\prod_{k=1}^{j}\frac{g_k}{f_k}},& i=1,2,\cdots,N-1\\
\end{array}
\end{equation}
In the limit of weak selection, the 1/3 law can be obtained. If $A$
and $B$ are strict Nash equilibria and the unstable equilibrium
occurs at a frequency of $A$ which is less than 1/3, then selection
favors replacement of $B$ by $A$. Moreover, stochastic evolution of
finite populations need not choose the strict Nash equilibrium and
can therefore favor cooperation over defection \cite{17}. \\
Obviously, the characteristic timescales also play a crucial role in
the evolutionary dynamics. Sometimes, although the mutant could
invade the entire population, it takes such a long time that the
population typically consists of coexisting strategies \cite{18}. It
can be shown that a single mutant following strategy $A$ fixates in
the same average time as a single $B$ individual does in a given
game, although the fixation probability for the two strategies are
different \cite{19}. Furthermore, if the population size is
appropriate for the fixation of the cooperative strategy, then this
fixation will be fast \cite{20}. Besides the standard Moran process,
Wright-Fisher model and pairwise comparison using Fermi function are
brought into the analysis of the evolutionary game in finite
populations \cite{21,22}. Moreover, spatial structure effects can
not be ignored in the real world. Much results reveal that a proper
spatial structure could enhance the fixation probability
\cite{23,24}.\\
Most results state the situation with two strategies. But in
reality, there may be many strategies in a game. Furthermore, in
coevolution of three strategies, how and why a single $A$ individual
could invade a finite population of $B$ and $C$ individuals, what
kinds of strategists would be washed out by the natural selection,
and how cooperation could emerge in finite populations are unclear.
Motivated by these, here we study the evolutionary game of finite
populations with three strategies. This paper is organized as
follows. Using the stochastic processes theory, we formulate the
evolutionary game dynamics in finite populations as a system of
linear equations in Section 2. The variable of these equations is
fixation probability $x(i,j)$, which represents the probability that
$i$ individuals using strategy $A$ could dominate a population in
which $j$ of them follow strategy $B$ and $N-i-j$ follow strategy
$C$. Two probabilities, the {\em global} and the {\em local}
fixation probability, act crucial roles in the evolutionary game
dynamics with three strategies. If the {\em global} fixation
probability of a single $A$ individual exceeds the neutral fixation
probability $1/N$, the selection favors $A$ replacing $B$ or $C$ no
matter what the initial ratio of $B$ to $C$ is. Similarly, if the
{\em local} fixation probability exceeds the neutral one $1/N$, the
selection favors $A$ replacing $B$ or $C$ only in some appropriate
initial ratios of $B$ to $C$. In Section 3, some numeric
computations of evolutionary game with AllC, AllD and TFT are
adopted to investigate the emergence of cooperation in some
specified situations. For weak selection and sufficiently large size
$N$, we find a condition in terms of the number of rounds $n$ and
the ratio $r$ of cost to benefit, under which the selection favors
only one TFT replacing AllC or AllD individuals. Furthermore, the
condition under which a single TFT could invade the entire
population is also obtained. Finally, the
results are summarized and discussed in Section 4. \\

\section{Model}
\label{} 
\begin{figure}[thpb]
\begin{center}
\includegraphics*[width=5cm]{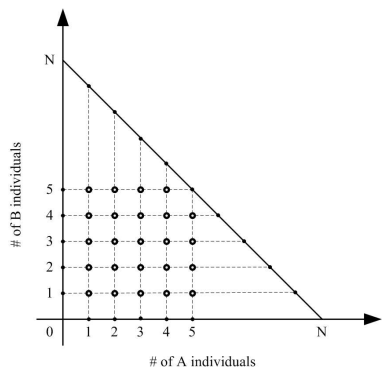}
\caption{The point $(i,j)$ marked by full black dot means the
boundary condition of the equations, while this marked by empty dot
means the unknown of the equations which could be formulated by the
boundary conditions.} \label{f1}
\end{center}
\end{figure}

Let us consider a well-mixed population of constant and
finite $N$ individuals . Suppose the strategy set in our model is
$A$, $B$ and $C$. The payoff matrix of the three strategies is
\[
\begin{array}{ccccccc}
  & & A&  & B& & C\\
A& &  a&  & b&  & p\\
B& &  c&  & d&  & q\\
C& &  m & & n&  & l
\end{array}
\]
The fitness of individuals using $A$, $B$ and $C$ is respectively as
follows:
\begin{equation}\label{e3}
\begin{array}{c}
f_{i,j}=1-w+w[a(i-1)+bj+p(N-i-j)]/(N-1)\\
g_{i,j}=1-w+w[ci+d(j-1)+q(N-i-j)]/(N-1) \\
h_{i,j}=1-w+w[mi+nj+l(N-i-j-1)]/(N-1)\\
\end{array}
\end{equation}
Here $i$ denotes the number of individuals using strategy $A$, $j$
denotes the number of those using strategy $B$, and there are
$N-i-j$ players using strategy $C$. The balance between selection
and drift can be described by a frequency-dependent Moran process.
At each time step, the number of $A$ individuals increases by one
corresponding to two situations. One is eliminating a $B$ individual
whereas the number of $C$ players keeps unchanged. The other is
eliminating a $C$ individual whereas the number of $B$ players keeps
unchanged. The transition probabilities can be formulated as:
\begin{equation}\label{e4}
\begin{array}{ccl}
 P_{i,i+1}^{j,j}&=&\frac{\displaystyle if_{i,j}}{\displaystyle if_{i,j}+jg_{i,j}+(N-i-j)h_{i,j}}\frac{\displaystyle N-i-j}{\displaystyle N} \\
 P_{i,i+1}^{j,j-1}&=&\frac{\displaystyle if_{i,j}}{\displaystyle if_{i,j}+jg_{i,j}+(N-i-j)h_{i,j}}\frac{\displaystyle i}{\displaystyle N} \\
 P_{i,i-1}^{j,j}&=&\frac{\displaystyle (N-i-j)h_{i,j}}{\displaystyle if_{i,j}+jg_{i,j}+(N-i-j)h_{i,j}}\frac{\displaystyle i}{\displaystyle N} \\
 P_{i,i-1}^{j,j+1}&=&\frac{\displaystyle jg_{i,j}}{\displaystyle if_{i,j}+jg_{i,j}+(N-i-j)h_{i,j}}\frac{\displaystyle i}{\displaystyle N} \\
 P_{i,i}^{j,j+1}&=&\frac{\displaystyle jg_{i,j}}{\displaystyle if_{i,j}+jg_{i,j}+(N-i-j)h_{i,j}}\frac{\displaystyle (N-i-j)}{\displaystyle N} \\
 P_{i,i}^{j,j-1}&=&\frac{\displaystyle (N-i-j)h_{i,j}}{\displaystyle if_{i,j}+jg_{i,j}+(N-i-j)h_{i,j}}\frac{\displaystyle j}{\displaystyle N} \\
 P_{i,i}^{j,j}&=& 1-P_{i,i}^{j,j+1}-P_{i,i}^{j,j-1}-P_{i,i-1}^{j,j+1}-P_{i,i-1}^{j,j}-P_{i,i+1}^{j,j-1}-P_{i,i+1}^{j,j}\\
\end{array}
\end{equation}
Here $P_{i,i+1}^{j,j}$ is the transition probability from the state
of $i$ $A$, $j$ $B$ and $N-i-j$ $C$ individuals to that of $i+1$
$A$, $j$ $B$ and $N-i-j-1$ $C$ individuals. Let $x(i,j)$ denotes the
fixation probability that $i$ $A$ individuals could invade the
population of $j$ $B$ and $N-i-j$ $C$ individuals. We have the
recursive relation:
\begin{equation}\label{e5}
\begin{array}{ccl}
x(i,j)&=&x(i+1,j)P_{i,i+1}^{j,j}+x(i+1,j-1)P_{i,i+1}^{j,j-1} \\
&+&x(i-1,j)P_{i,i-1}^{j,j}+x(i-1,j+1)P_{i,i-1}^{j,j+1} \\
&+&x(i,j+1)P_{i,i}^{j,j+1}+x(i,j-1)P_{i,i}^{j,j-1}+x(i,j)P_{i,i}^{j,j}\\
\end{array}
\end{equation}
Researchers reported that in a well-mixed environment, two of the
initial three kinds of strategists would go extinct after some
finite time, while coexistence of the populations was never observed
\cite{25}. In finite populations, no matter how many kinds of
individuals initially, only one type of strategists can survive in
the evolutionary game eventually. Hence, the fixation probabilities
of the game with two strategies can be viewed as special boundary
conditions of our model. There are three types of boundary
conditions:\\
(1)obviously, $x(0,j)=0$, $j=0,1,\cdots,N$;\\
(2)$x(i,0)=\tilde{x}(i)$, $i=0,1,\cdots,N$, here $\tilde{x}(i)$ is
the fixation probability that $i$ $A$ individuals could take over
the population of $N-i$ $C$ and no $B$ players; \\
(3)similarly, $x(i,N-i)=\bar{x}(i)$, $i=0,1,\cdots,N$, here
$\bar{x}(i)$ means the fixation probability that $i$ $A$ individuals
could invade the population of $N-i$ $B$ and no $C$ players.\\
Note that $\bar{x}(i)$ can be formulated as Eq. \ref{e2}. Similarly,
$\tilde{x}(i)$ is written as
$$\tilde{x}(i)=\frac{\displaystyle 1+\sum_{j=1}^{i-1}\prod_{k=1}^{j}\frac{h_k}{f'_k}}{\displaystyle 1+\sum_{j=1}^{N-1}\prod_{k=1}^{j}\frac{h_k}{f'_k}}$$
here $i=1,2,\cdots,N-1$,  $f'_i=1-w+w[a(i-1)+p(N-i)]/(N-1)$,
 $h_i=1-w+w[mi+l(N-i-1)]/(N-1)$, and the corresponding boundary conditions
are $\tilde{x}(0)=0$,
 $\tilde{x}(N)=1$.\\
The relationship among the solutions of the system of equations can
be depicted by Fig. \ref{f1}. The point $(i,j)$ marked by full black
dot denotes the boundary condition of the equations, while this
marked by empty dot denotes the unknown of the equations. Thus in
what follows, the unknown element $x(i,j)$ of our interest is
discussed.

\begin{figure}[thpb]
\begin{center}
      \includegraphics[width=6.5cm]{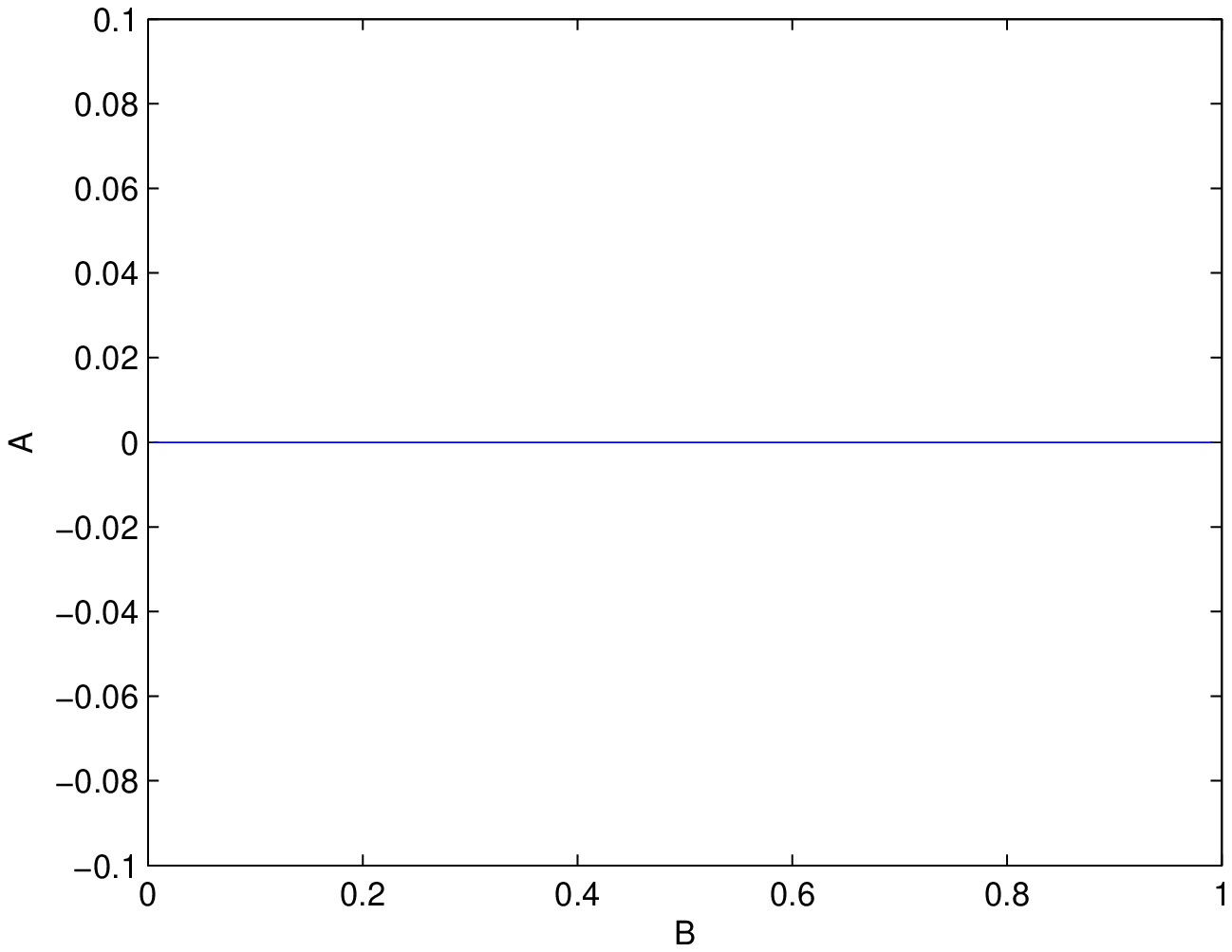}
      \includegraphics[width=6.5cm]{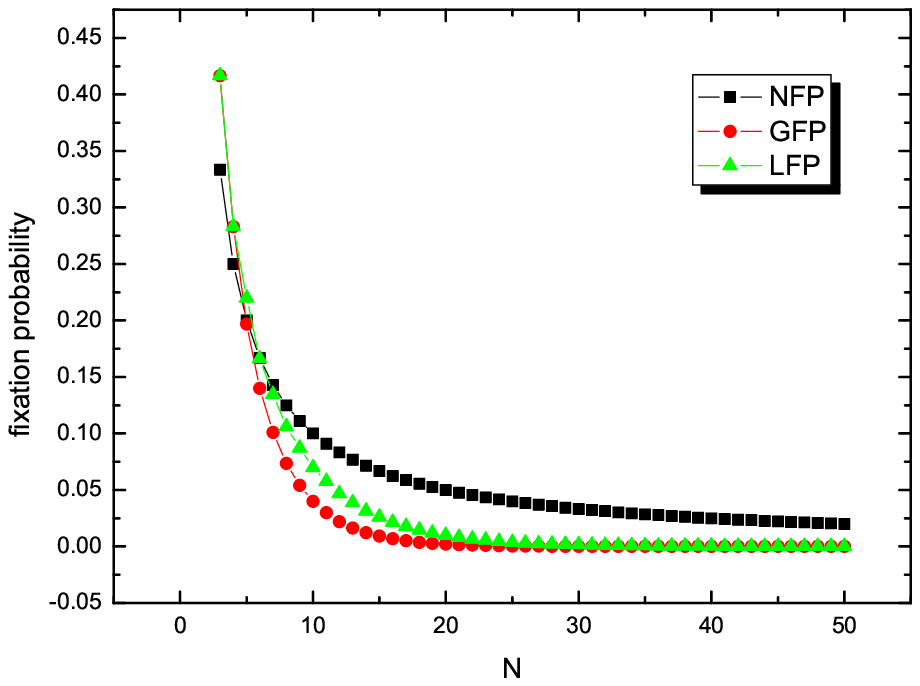}
      \includegraphics[width=6.5cm]{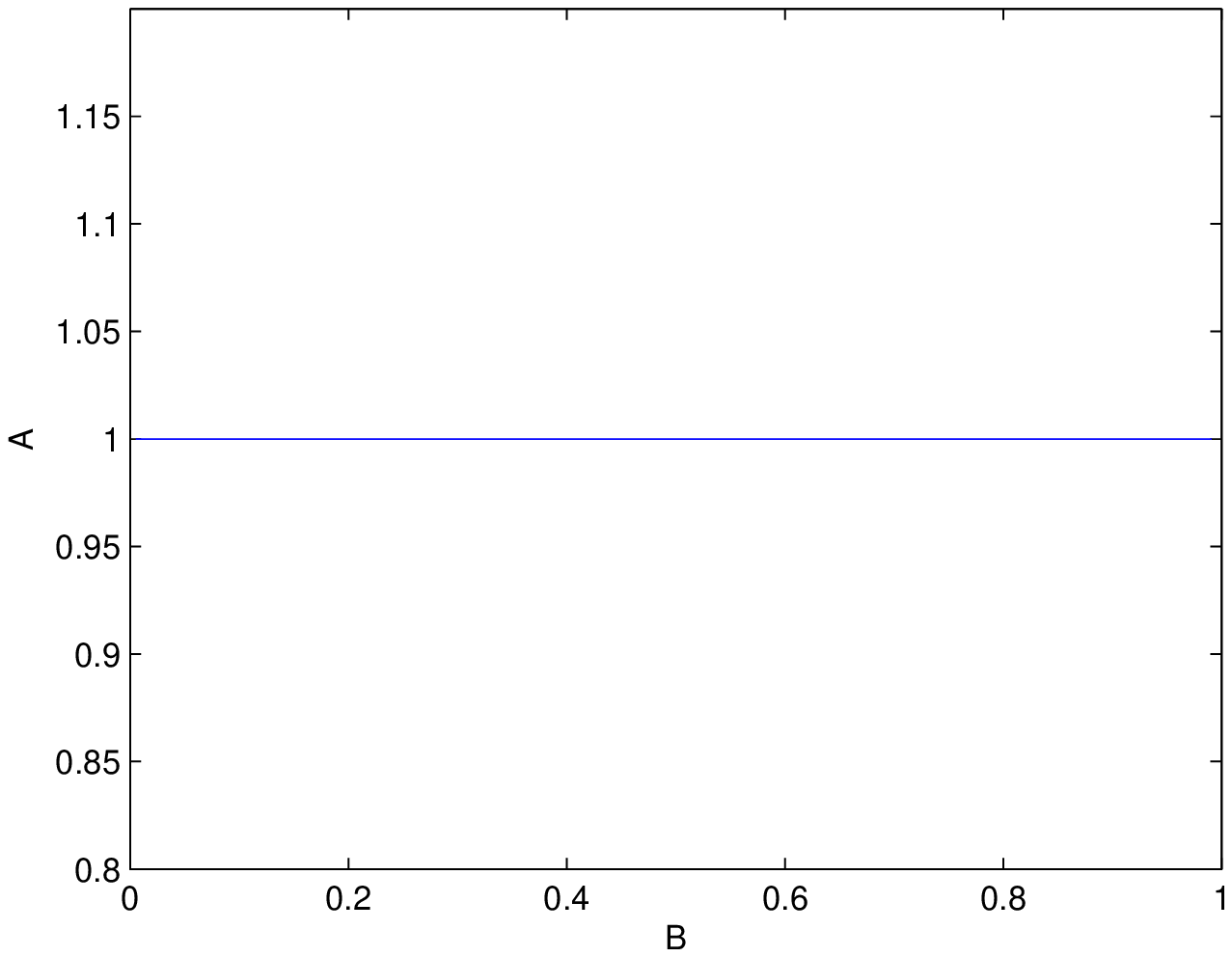}
      \includegraphics[width=6.5cm]{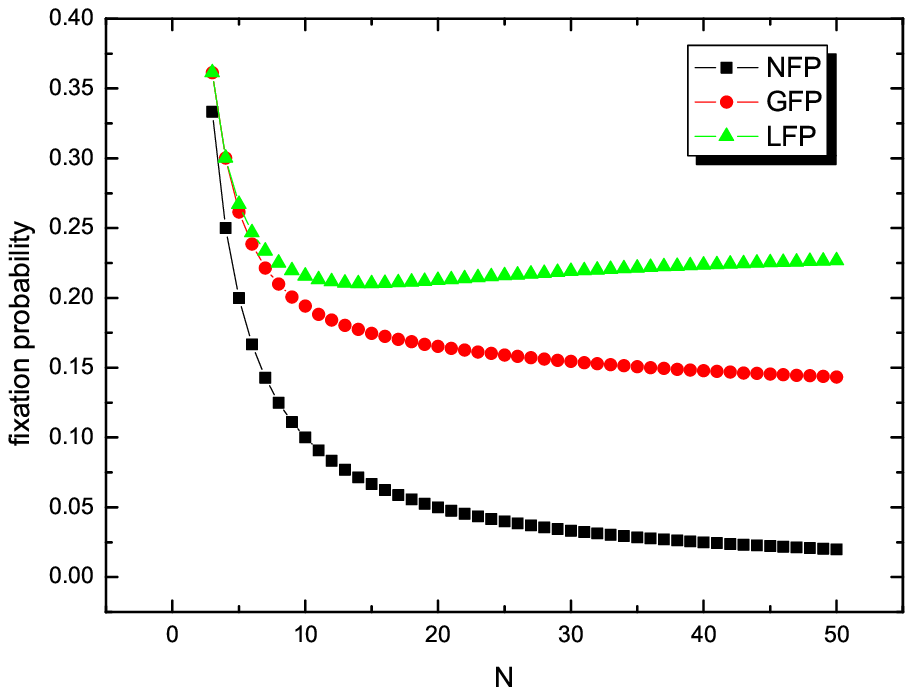}
      \includegraphics[width=6.5cm]{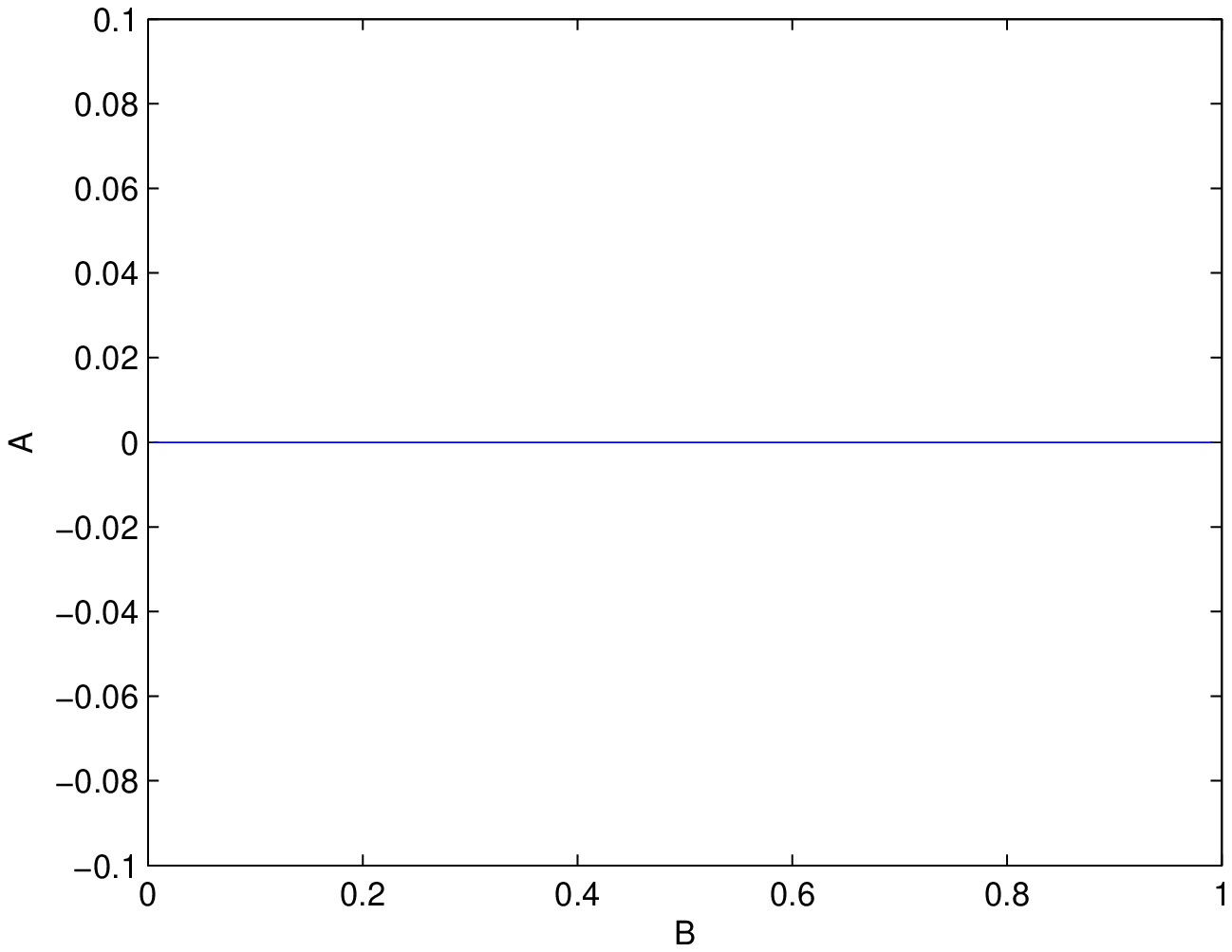}
      \includegraphics[width=6.5cm]{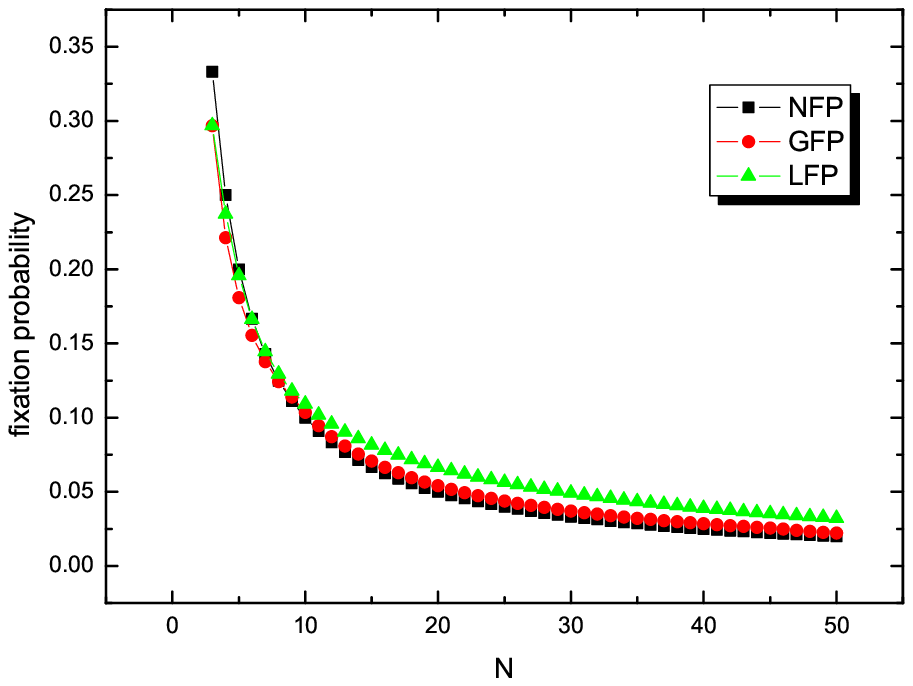}
      \caption{The comparisons between different situations corresponding to infinite and finite populations respectively. In the left column figures, the $x-$axis denotes the percentage of
$B$ individuals initially, meanwhile, the $y-$axis denotes the
percentage of $A$ individuals after 1000 steps. In the right column
figures, the $x-$axis shows the size of the population, and the
$y-$axis shows the fixation probability of a single $A$. Let NFP,
LFP and GFP represent the neutral fixation probability $1/N$, the
{\em local} and the {\em global} fixation probability respectively.
}
      \label{f2}
      \end{center}
\end{figure}

Eq. \ref{e5} can be transformed to Eq. \ref{e6} which is a system of
linear equations in $(N-1)(N-2)/2$ variables.
\begin{equation}\label{e6}
\begin{array}{lll}
& &[i(N-i)f_{i,j}+j(N-j)g_{i,j}+(i+j)(N-i-j)h_{i,j}]x(i,j)\\
&=&i(N-i-j)h_{i,j}x(i-1,j)+ijg_{i,j}x(i-1,j+1)\\
&+&i(N-i-j)f_{i,j}x(i+1,j)+ijf_{i,j}x(i+1,j-1)\\
&+&j(N-i-j)g_{i,j}x(i,j+1)+j(N-i-j)h_{i,j}x(i,j-1)\\
\end{array}
\end{equation}
Where $i=1,2,\cdots,N-2$, $j=1,2,\cdots,N-i-1$. Accordingly, Eq.
\ref{e5} can also be simplified to $A\vec{x}=\vec{b}$, where
$\vec{x}$ is a vector
$(x(1,1),x(1,2),\cdots,x(1,N-2),x(2,1),x(2,2),\cdots,x(2,N-3),\cdots,x(i,1),x(i,2),\cdots,x(i,N-i-1),\cdots,x(N-2,1))^{T}$,
$A$ is the corresponding coefficient
$(N-1)(N-2)/2\times(N-1)(N-2)/2$ dimensional matrix, $\vec{b}$ is
the corresponding vector composed of these boundary conditions.
Matrix $A$ can be written as follows:
\begin{equation}
\left(
\begin{array}{ccccccc}
A_{N-2,N-2}& A_{N-2,N-3}& 0& 0& \cdots& 0& 0\\
A_{N-3,N-2}& A_{N-3,N-3}& A_{N-3,N-4}& 0& \cdots& 0& 0\\
 0& A_{N-4,N-3}& A_{N-4,N-4}& A_{N-4,N-5}& \cdots& 0& 0\\
 0& 0& A_{N-5,N-4}& A_{N-5,N-5}& A_{N-5,N-6}& \cdots& 0\\
\vdots& &\vdots& &\vdots& &\vdots\\
0& 0& \cdots& 0& A_{2,3}& A_{2,2}& A_{2,1}\\
0& 0& \cdots& 0& 0& A_{1,2}& A_{1,1}\\
\end{array}
\right)
\end{equation}
Here $A_{N-i,N-j}$ states a $(N-i)\times(N-j)$ block of matrix
$A$.\\
For fixed $i$, in the sub-vector
$(x(i,1),x(i,2),\cdots,x(i,N-i-1))^T$, there exist a maximal
probability $x_{max}(i)$ and a minimal probability $x_{min}(i)$.
They both play significant roles in the evolution process. Note that
in evolutionary dynamics, we have $x_{max}(1)\leq
x_{max}(2)\leq\cdots\leq x_{max}(N-2)$, $x_{min}(1)\leq
x_{min}(2)\leq\cdots\leq x_{min}(N-2)$. If a single $A$ individual
can be favored to replace $B$ or $C$, more than one $A$ individuals
are more likely to replace $B$ or $C$. Thus we only focus on the
fixation probabilities $x(1,j)$. For a neutral mutant, the fixation
probability is $1/N$. When $x(1,j)>1/N$, natural selection favors
$A$ replacing $B$ or $C$. We define $x_{min}(i)$ as {\em global}
fixation probability and $x_{max}(i)$ as {\em local} fixation
probability of $i$ $A$ individuals. If $x_{min}(1)>1/N$, natural
selection favors $A$ replacing $B$ or $C$ no matter what the ratio
of $B$ to $C$ is , that is {\em global}. Similarly, if
$x_{max}(1)>1/N$, the selection favors $A$ replacing $B$ or $C$ in
some proper ratios of $B$ to $C$, not any ratios, that is {\em
local}. Accordingly, there may be
three situations:\\
 \\
$(1)$ if $x_{max}(1)<1/N$, natural selection never favors $A$
replacing $B$ and $C$;\\
 \\
$(2)$ if $x_{min}(1)>1/N$, natural selection always favors $A$
replacing $B$ or $C$ whatever the ratio of $B$ to $C$ is. It is
likely that $A$ could invade the population; \\
 \\
$(3)$ if $x_{min}(1)<1/N$ and $x_{max}(1)>1/N$, natural selection
favors $A$ replacing $B$ or $C$ in some proper ratios of $B$ to $C$.
Thus the fixation of $A$ is possible under some suitable
conditions.\\
 \\
Let us compare the evolutionary game dynamics in infinite
populations with that in finite populations. The finite size effects
bring stochastic factor to the evolution. $A$ individuals which are
eliminated in infinite populations may be favored by natural
selection replacing $B$ or $C$ individuals. In Fig. \ref{f2}, the
left column shows evolutionary game with a small frequency of $A$
individuals initially in infinite populations, and the right column
shows that with a single $A$ individual at first in finite
populations. The first row shows situations with payoff matrix $a=2,
b=4, p=4, c=3, d=5, q=1, m=3, n=1, l=5$, the size $N=50, w=0.1$. $A$
individuals will disappear in infinite populations no matter how
many $B$  and $C$ individuals are initially. While in finite
populations, the selection won't favor $A$ replacing $B$ or $C$. The
second row shows the situations with payoff matrix $a=5, b=4, p=4,
c=4, d=3, q=4, m=4, n=4, l=3$, the size $N=50, w=0.1$. $A$
individuals will always invade the population of $B$ and $C$
individuals in infinite populations. And in finite populations,
natural selection will all the time favor $A$ replacing $B$ or $C$
no matter what the ratio of $B$ to $C$ is. The third row shows
situations with payoff matrix $a=5 ,b=5 ,p=1.6 ,c=5 ,d=5 ,q=0 ,m=3
,n=7 ,l=2, N=50, w=0.1$.  $A$ individuals will disappear in infinite
populations no matter how many $B$ and $C$ individuals are
initially. Furthermore, from plenty of computations, we find that
$A$ individuals will monotonously decrease to zero in infinite
situation. Whereas sometimes $A$ will tend to replace  $B$ or $C$ in finite populations.\\

\begin{figure}[thpb]
\begin{center}
      \includegraphics[width=8.5cm]{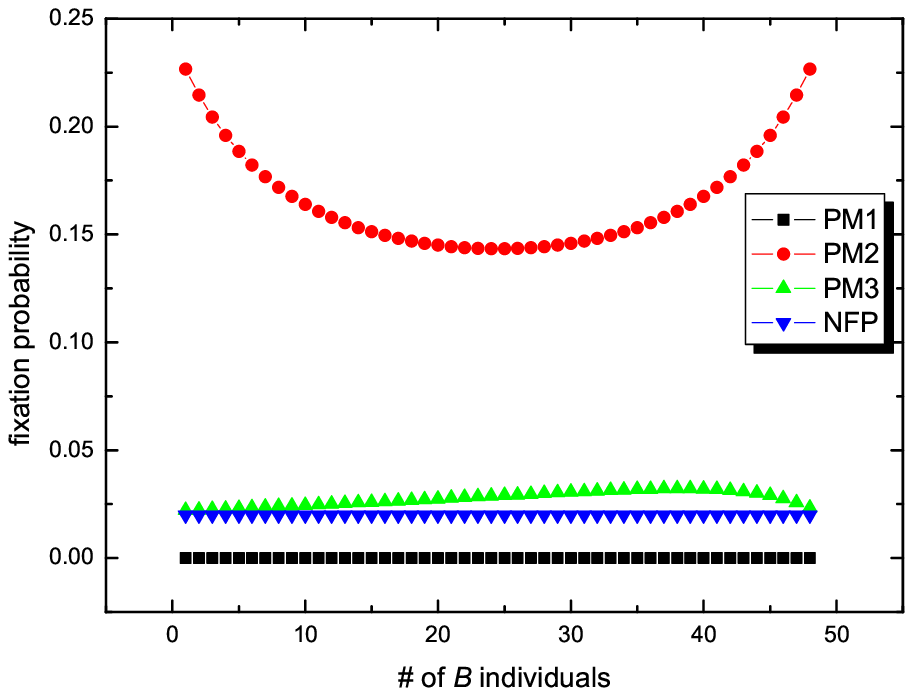}
      \caption{The fixation probability of only one
$A$ individual as a function of the number of $B$ individuals. Where
$N=50$, NFP is the neutral fixation probability, PM$i$ means the
situation with payoff matrix $i$ in Fig. \ref{f2}, $i=1,2,3$. The
fixation probability of only one $A$ individual is not always
monotonic function of the number of $B$ individuals. The {\em
global} and {\em local} fixation probability may be not at the end
points of the curve.}
      \label{f3}
      \end{center}
\end{figure}

For the three payoff matrixes in Fig. \ref{f2}, it is clear that the
fixation probability of $A$ individual is not always monotonic
function of the number of $B$ individuals (see Fig. \ref{f3}). Thus,
the {\em global} and {\em local} fixation probability of $A$ may
nontrivially occur
at intermediate ratio of B to C, not always end points of the ratio range of B to C. \\

\section{AllC-AllD-TFT}
\label{}

\begin{figure}[thpb]
\begin{center}
      \includegraphics[width=8.5cm]{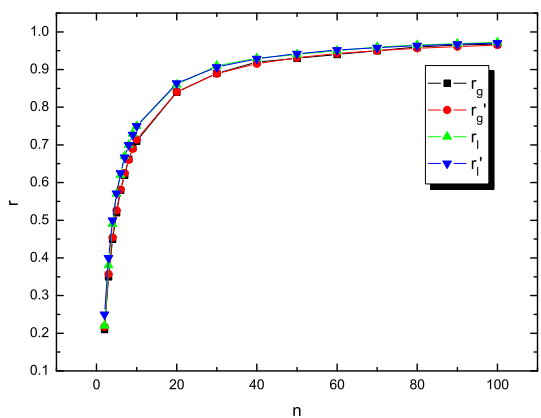}
      \caption{The {\em global} and {\em local}
barrier ratios $r_{g}$ and $r_{l}$ as a function of the number of
rounds $n$.
       Here $N=50, w=0.1$. $r_{g}'$ represents the fitting  curve of $r_{g}$, $r_{l}'$ represents the fitting curve of $r_{l}$.
       We can fit $r_{l}$
and $r_{g}$ as $r_{l}=\frac{n-1}{n+2}$, and
$r_{g}=\frac{n-1}{n+2.6}$.}
      \label{f4}
      \end{center}
\end{figure}

Let us consider a very interesting and famous repeated game with
three strategies AllC (cooperate all the time), AllD (defect all the
time) and TFT (tit-for-tat). TFT is an adaptive cooperative strategy
which is one of the most successful strategies proved by
experiments. Individuals using TFT cooperate in the first round
generally, and then do whatever the opponents did in the previous
round. The number of rounds $n$, by definition, can be $1, 2,
\cdots, \infty$. If the rounds are infinite, this game is out of our
consideration. In one-round repeated Prisoners' Dilemma game, a
cooperator can obtain a benefit of $b$ or $0$ if it meets a
cooperator or defector, meanwhile, it must cost $c$ whomever its
opponent is; a defector can obtain a benefit of $b$ or $0$ if it
meets a cooperator or defector, but costs nothing in the whole
process. We bring the ratio of cost to benefit, by definition, $r$
($r=c/b$$\in$[0,1]), into our game, the payoff matrix between
cooperator and defector can be simplified as follows:
\[
\begin{array}{ccc}
  & C& D\\
C&  1& 0\\
D&  1+r&  r
\end{array}
\]
Thus the payoff matrix of TFT, AllC and AllD with $n$ rounds is
\[
\begin{array}{cccc}
  & TFT& AllC& AllD\\
TFT&  n& n& (n-1)r\\
AllC&  n&  n& 0\\
AllD&  1+nr&  n(1+r)& nr
\end{array}
\]
The pairwise comparison of the three strategies leads to the
following conclusions. \\
(1) AllC is dominated by AllD, which means it is best to play AllD
against both AllC and AllD; \\
(2) TFT is equal to AllC when TFT plays with AllC; \\
(3) If the average number of rounds exceeds a minimum value,
$n>1/(1-r)$, then TFT and AllD are bistable. \\
Suppose that a single individual using strategy TFT is brought in
the population in which some individuals adopt strategy AllC and the
others use strategy AllD originally. Provided that the number of
rounds is finite and greater than $1/(1-r)$, the strategies TFT and
AllD are both strict Nash equilibrium and evolutionary stable
strategies (ESS) \cite{2}. If $n=1$, TFT becomes strategy $C$ which
is out of our discussion. Thus we need finite $n\gg2$. For a fixed
number of individuals and a value of the rounds, we find that there
is a barrier of $r$ which can determine whether or not the selection
favors TFT replacing AllC or AllD. The barrier also has two types:
one is $r_{l}$, which represents the barrier of {\em local}
situation; the other is $r_{g}$, which represents that of {\em
global} situation. If $r<r_{l}$, the natural selection favors TFT
replacing AllC or AllD {\em locally}, whereas if $r>r_{l}$, TFT
tends to be washed out by selection. The results about $r_{g}$ are
similar. Under the condition of weak selection, for sufficiently
large population size $N$ and large number of rounds $n$, the
barrier ratio $r$ as the function of $n$ is approximately followed
by $(n-1)/(n+\theta)$, here $\theta$ is a parameter dependent with
$N$. In Fig. \ref{f4}, we can fit $r_{l}$ and $r_{g}$ as
$r_{l}=(n-1)/(n+2)$, and $r_{g}=(n-1)/(n+2.6)$. In general, for
$w\ll1$, sufficiently large $N$ and large $n$, from abundant numeric
computations, we obtain $r_{l}=(n-1)/(n+2)$ and
$r_{g}=(n-1)/(n+\theta_{g})$, here $\theta_{g}$ is also a parameter
dependent with $N$ and $\theta_{g}>2$. Thereforce, $r_{l}> r_{g}$.
In other words, the ratio $r$ which could lead the selection to
favor TFT replacing AllC or AllD
{\em globally} can also induce {\em local} replacement, but not vice versa.\\
Deterministic replicator dynamics with three strategies in infinite
populations admits two interior equilibria at frequency of TFT given
by $x^{*}=r/(n-1)(1-r)$ and $x^{**}=nr/(n-1)$. For {\em local
situation}, substitude $r<(n-1)/(n+2)$ into $x^{*}$ and $x^{**}$, we
get $x^{*}<1/3$ and $x^{**}<n/(n+2)$; for global situation,
substitude $r<(n-1)/(n+\theta_{g})$ into $x^{*}$ and $x^{**}$, we
obtain $x^{*}<1/(1+\theta_{g})$ and $x^{**}<n/(n+\theta_{g})$. If
the frequency of TFT at the equilibrium is $x^{*}<1/3$ or
$x^{**}<n/(n+2)$ in infinite populations, it will be favored
replacing AllC or AllD {\em locally} in finite situation; if the
frequency of TFT at the equilibrium of infinite situation is
$x^{*}<1/(1+\theta_{g})$ or $x^{**}<n/(n+\theta_{g})$, it will tend
to replace AllC or AllD {\em globally} by TFT in finite
populations.   \\
The  1/3 law proposed by Nowak in \cite{12} is still valid in our
case, that is, the selection favors TFT replacing AllC or AllD in
finite populations, if its frequency at the equilibrium is
$x^{*}<1/3$ in infinite populations. However, when there are three
strategies TFT, AllC and AllD, in which TFT's frequency at one
equilibrium is $x^{*}$, the corresponding frequency of AllC is zero.
Thus in this situation, our results to some extent validate the
conjecture in which AllC is eliminated by natural selection so
quickly that the effect of AllC can be neglected in finite
populations. And then the evolutionary game dynamics with the two
left strategies TFT and AllD, is equivalent to the situation of the
situation with these two strategies initially. Nevertheless not any
size of AllC individuals could be wiped out quickly, their effects
can not be ignored in the dynamics. Hence $x^{*}<1/3$ can only
determine the replacement {\em locally} (in some certain
circumstances). As for {\em global} fixation situation (the fixation
is certain for any ratio of AllC to AllD), we have
$x^{*}<1/(1+\theta_g)<1/3$ for $\theta_g>2$. The conditions that
natural selection favors {\em global} replacement of AllC or AllD by
TFT are more intensified than
those of {\em local} situation.\\
Let us discuss the other equilibrium. The $n/(n+\theta)$ is a
monotony increasing function of $n$. This $n/(n+\theta)$ approaches
one for increasing $n$. That is to say, when the number of rounds
increases, the condition $x^{**}<n/(n+\theta)$ can be satisfied with
higher probability, and TFT may have more opportunities to replace
AllC or AllD {\em locally} and {\em globally}. In the standard
evolutionary model of the finitely repeated Prisoner's Dilemma, TFT
can not invade AllD. But interestingly, we find that for
intermediate $n$, if $x^{**}<n/(n+\theta)$, nature selection favors
TFT replacing AllC or AllD as the fixation probability of a single
TFT ($\rho_{TFT}$) is larger than that of a single AllC
($\rho_{AllC}$) or AllD ($\rho_{AllD}$). Actually, in this case,
$\rho_{TFT}>1/N>max(\rho_{AllD},\rho_{AllC})$. Therefore, a single
TFT is likely to invade the entire population consisting of AllC and
AllD finally. In this case, cooperation tends to emerge in the
evolution process. And yet, for large limit $n$, the situation is
out of our consideration due to its extraordinary intricacy.
However, as $n$ increasing to infinite, the probability that the
selection favors TFT taking over the whole population also
approaches one. Accordingly, the fixation of cooperation is enhanced
in finite populations. It is because that when TFT meets AllD, its
loss in the first round can be diluted by many rounds games. In this
case, the total fitness of TFT is almost the same as that of AllD
and they are a pair of nip and tuck opponents. But TFT receives more
payoff than AllD when they both play with TFT. To sum up, TFT is
superior to AllD for limit large rounds because of its adaption. As
a result of this property of TFT, natural selection mostly prefers
to choose TFT to reproduce offspring, and then TFT is most likely to
dominate the population at last. Therefore, cooperation has more
opportunities to win in finite populations contrasting against
infinite situation. \\
\section{Conclusion}
\label{}

We have proposed a model of evolutionary game dynamics with three
strategies in finite populations. It can be characterized by a
frequency-dependent Moran process which could be stated by a system
of linear equations. By the comparative study of evolution in finite
and infinite populations, we shew that a single $A$ individual which
can not invade infinite populations may have an opportunity to
replace $B$ or $C$ in finite situation. In other words, a single $A$
individual could be eliminated by selection with smaller probability
in finite populations than situation in infinite populations. In
addition, a famous game with AllC, AllD, and TFT is adopted to
illuminate our results by numeric computations. Furthermore, under
the condition of weak selection, for sufficiently large population
size $N$ and appropriate number of rounds $n$, a single TFT could
invade the population composed of AllC and AllD with high
probability almost one. In this situation, the emergence of
cooperation is attributed to the finite population size effects. Our
results may help understand the
coevolution of multi-species and diversity of natural world.\\

\section*{Acknowledgement}
We are grateful to Xiaojie Chen and Bin Wu for helpful discussions
and comments. This work was supported by National Natural Science
Foundation of China (NSFC) under grant Nos. 60674050 and 60528007,
National 973 Program (Grant No.2002CB312200), National 863 Program
(Grant No.2006AA04Z258) and 11-5 project (Grant No. A2120061303).



\end{document}